\documentclass[preprint2]{aastex62}
\usepackage{textcomp}
\usepackage{gensymb}
\usepackage{subfigure}
\usepackage{amsmath}

\shorttitle{Dust in the outskirts of M\,31 and M\,33}
\shortauthors{Zhang \& Yuan}

\DeclareGraphicsExtensions{.eps,.pdf,.jpg,.bmp}

\begin{document}
\title{Detections of dust in the outskirts of M\,31 and M\,33}
\author[0000-0002-0702-7551]{Zhang Ruoyi}
\author[0000-0002-0702-7551]{Yuan Haibo}
\affil{Department of Astronomy, Beijing Normal University\\ No.19, Xinjiekouwai St, Haidian District, Beijing, 100875, P.R.China;  yuanhb@bnu.edu.cn}

\begin{abstract}
M\,31 and M\,33 serve as ideal places to study distributions of dust in the outskirts of spiral galaxies.
In this letter, using about 0.2 million stars selected from the LAMOST data and combining precise photometry and parallaxes from the {\it Gaia} DR2,
we have constructed a two-dimensional foreground dust reddening map towards the M\,31 and M\,33 region ($111.2^\circ \le gl \le 136.2^\circ$, $-36.5^\circ \le gb \le -16.5^\circ$).
The map has a typical spatial resolution of about 12 arc-minute and precision of 0.01 mag. 
The complex structure of dust clouds towards the M\,31 is revealed.
By carefully removing the foreground extinction from the dust reddening map of Schlegel et al. (1998), we thus have obtained a residual map to study dust distributions in the outskirts of M\,31 and M\,33.
A large amount of dust is detected in the M\,31 halo out to a distance of over 100 kpc. Dust in the M\,31 disk is found to extend out to about 2.5 times its optical radius, whose distribution is 
consistent with either an exponential disk of scale length of 7.2 kpc or two disks with scale length of 11.1 kpc within its optical radius and 18.3 kpc beyond its optical radius. 
Dust in the disk of M\,33 is also found to extend out to about 2.5 times its optical radius, its distribution beyond one optical radius is consistent with an exponential disk of scale length of 5.6 kpc.
Our results provide new clues on the distributions and cycling of dust in galaxies.
\end{abstract}

\keywords{dust, extinction; galaxies: individual (M\,31, M\,33)
}

\section{INTRODUCTION}

Dust is ubiquitous in the universe, from the interstellar medium (ISM; Draine 2003), the circum-galactic medium (CGM; Tumlinson et al. 2017) to the intra-galactic medium (IGM; Shchekinov \& Nath 2011). 
Studies of distributions and properties of dust from the Milky Way to extra-galaxies have wide implications, including formation, transportation and destruction of dust, 
formation and evolution of galaxies, and precise reddening correction to reveal the intrinsic properties of astronomical objects.

It has been known for decades that galactic dust/gas disks extend beyond the dimensions inferred from the stellar disks in both the Galaxy and extra-galaxies. 
By comparing scale lengths in different photometric bands, dust scale lengths are also found to be much larger than the stellar ones, up to a factor of 2 (e.g., Casasola et al. 2017).
The presence of extended dust distributions in spiral galaxies is also supported by far-infrared observations of dust emission.
By modeling the COBE observations at 140$\mu$ and 240$\mu$, Davies et al. (1997) find a dust disk with a radial scale length 1.5 times the stellar for the Galaxy. 
The result is further confirmed by three-dimensional modeling of the Galactic dust distribution with the LAMOST data (Li et al. 2018). 
With similar techniques, evidence for an extended distribution of cold dust was also found for extra-galaxies (e.g., Hinz et al. 2006). 
By combining the signal of 110 spiral galaxies, Smith et al. (2016) report the direct detection of dust emission that extends out to at least twice the optical radius.
And they find that the distribution of dust is consistent with an exponential at all radii with a gradient of about $-$1.7 dex per $R_{25}$, corresponding to a dust scale length of 0.44 $R_{25}$.
Using an occulting galaxy technique, Holwerda et al. (2009) find a dusty disk much more extended than the starlight, with spiral lanes seen in extinction out to 1.5 $R_{25}$ radii.

Dust is also believed to exist in the CGM. 
Zaritsky (1994) reports a preliminary detection of a dusty halo, through the additional color excess of background objects in fields close to two galaxies, compared to that of field objects.
Based on the angular correlation between the reddening of about 85,000 quasars at z $>$ 1 and the positions of 24 million galaxies at z $\sim$ 0.3 detected in the SDSS (York et al. 2000), 
M{\'e}nard et al. (2010) have detected the presence of extended halos of dust from 20 kpc to several Mpc. 
Dust clouds in the Galactic halo have also been detected to a distance of about 30 kpc (Yuan et al. in preparation).

M\,31 and M\,33, as respectively the largest and 3rd largest galaxies in the Local Group, serve as ideal places not only to study the dust in external spiral galaxies with an extremely high spatial resolution (e.g., Draine et al. 2014), but also to study distributions of dust in the very outskirts of galaxies.
Tempel et al. (2010) have constructed a three-dimensional galaxy model with axisymmetric stellar populations and a dust disk to estimate extinction and the intrinsic luminosity and color 
distributions of M\,31, using the Spitzer far-IR maps to determine the dust distribution. 
The resulting scale length of the dust disk with a simple exponential law is 8.5 kpc, about 1.8 times the stellar scale length.
However, this work is only limited to about one optical radius. Dust distributions in the outskirts of M\,31 and M\,33 have not been investigated yet, due to the 
difficulties in separating its weak signal from the foreground Galactic dust, either in far-IR emission or in absorption.

Thanks to the scientific significance of M\,31 and M\,33 
and the good match between the size (and position) of M\,31 and the field of view (and site weather) of LAMOST (Cui et al. 2012), 
M\,31 and its vicinity region has been extensively targeted by the LAMOST during its commissioning phase, pilot and regular survey phases (Yuan et al. 2015). 
A large number of foreground stars have been well observed and their reddening values 
can be accurately estimated with the so-called star-pair technique (e.g., Yuan et al. 2013). 
In this work, we use accurate reddening estimates of 193,847 stars to construct a foreground dust reddening map towards the M\,31 and M\,33 region.
Then by comparing with the dust reddening map of Schlegel et al. (1998), we aim to explore how dust is distributed in the outskirts of the two galaxies. 
The paper is organized as follows. In Section\,2 we introduce our data and method used to construct the foreground reddening map.
The resulting  map and distributions of dust in the outskirts of M\,31 and M\,33 are presented in Sections\,3 and \,4. 
The results are discussed in Section\,5. We summarize in Section\,6.                                                                                                                     

The following general parameters of M\,31 and M\,33 are adopted in this work:
for M\,31, the right ascension is 10.68458$^\circ$ and the declination is 41.26875$^\circ$ (Jarrett et al. 2003), the optical radius is 1.58$^\circ$, 
the inclination angle is 77.8$^\circ$ (de Vaucouleurs et al. 1991), the major axis position angle is 38.0$^\circ$ 
(Chemin et al. 2009) and the distance is 785 kpc (McConnachie et al. 2005); 
For M\,33, the right ascension is 23.46208$^\circ$ and the declination is 30.65994$^\circ$ (Jarrett et al. 2003), the optical radius is 0.59$^\circ$, 
the inclination angle is 53.9$^\circ$ (de Vaucouleurs et al. 1991), the major axis position angle is 50.5$^\circ$ 
(Jarrett et al. 2003) and the distance is 809 kpc (McConnachie et al. 2005). 

\section{DATA AND METHOD}\label{sect:data}

\subsection{Data}

LAMOST spectroscopic data (Zhao et al. 2012; Deng et al. 2012; Liu et al. 2014) combined with 
{\it Gaia} DR2 (Gaia Collaboration et al. 2018) photometry and parallaxes are used to construct 
a foreground dust reddening map towards the M\,31 and M\,33 region in this work.

We first select spectroscopic data from the LAMOST DR5, which has delivered more than 8 million stellar spectra with spectral 
resolution $R = 1800$ and limiting magnitude of r $\sim$\,17.8 mag (Deng et al. 2012; Liu et al. 2014). Stellar effective temperatures, surface gravities, and metallicities are 
derived by the LAMOST Stellar Parameter Pipeline (LASP; Wu et al. 2011), with the precision of about 110\,K, 0.2\,dex, and 0.15\,dex, respectively (Luo et al. 2015). 
Additional targets in the 2nd value-added catalog (Xiang et al. 2017) of the LAMOST Spectroscopic Survey of the Galactic Anti-centre (LSS-GAC; Liu et al. 2014, Yuan et al. 2015) are also 
used. As the LSS-GAC DR2 value-added catalog contains targets that were observed during the commissioning and testing time and not included in the LAMOST official data releases.  
Basic stellar parameters in the LSS-GAC value-added catalogs are derived with the LAMOST Stellar Parameter Pipeline at Peking University (LSP3; Xiang et al. 2015, 2017). 

The following criteria are used to select sample stars: 1) $111.2^\circ \le glon \le 136.2^\circ$ and $-36.5^\circ \le glat \le -16.5^\circ$; 
2) SNR $\ge$ 10; 3) positive {\it Gaia} DR2 parallaxes; and 4) $ 1.0+0.015(BP-RP)^2<$ phot\_bp\_rp\_excess\_factor $<1.3+0.06(BP-RP)^2 $ 
to exclude stars whose BP/RP photometry is probably contaminated from nearby sources (Evans et al. 2018). In cases of duplicated targets, the one of the higher/highest SNR is used. 
In total, about 160,000 stars from the LAMOST DR5 and 40,000 stars from the LSS-GAC DR2 are selected. 

\subsection{Estimates of reddening for individual stars}

A straightforward star-pair method is adopted to obtain reddening values of the selected stars. The method assumes that stars of the same stellar atmospheric parameters have the same intrinsic colors. Thus, 
the intrinsic colors of a reddened star can be derived from its control pairs/counterparts of the same atmospheric parameters ($\Delta T_{\rm eff} < T_{\rm eff} *{(0.00003*T_{\rm eff})}^2\,K, \Delta log g < 0.5\,dex, 
\Delta [Fe/H] < 0.3\,dex$) that suffer from either nil or well-known extinction (Yuan et al. 2013). 
For the control stars, their reddening values are from the SFD98 map and required to be smaller than 0.02 mag, 
their distances to the Galactic plane are required to be larger than 1 kpc to ensure that they are far above the Galactic dust disk.
Given the high precision of {\it Gaia} DR2 photometry, reddening values of $E(BP-RP)$ are computed in this work, using the same star-pair algorithm of Yuan et al. (2015; See their Section\,5 for more details). 

To make our Galactic foreground reddening map in the M\,31 and M\,33 region directly comparable to the SFD98 map, values of $E(BP-RP)$ are converted to $E(B-V)$ 
via the following temperature and reddening dependent reddening coefficient $R_{BP-RP}$ (Sun et al. in preparation):
\begin{multline} 
R_{BP-RP} = 1.060 - 0.253 \times E(B-V)_{SFD} + \\
0.264 \times E(B-V)_{SFD}^2 + 5.535 \times 10^{-6} \times T_{eff} - 4.505\\
\times 10^{-5} \times T_{eff} \times E(B-V)_{SFD} + 6.639 \times 10^{-9} \times T_{eff}^2 
\end{multline}
The formula is empirically determined from a large sample of LAMOST DR5 stars, with their values of 
$E(BP-RP)$ estimated using the same star-pair algorithm and values of $E(B-V)$ from the SFD98 map.
The hotter stars are, the smaller reddening stars suffer, the larger $R_{BP-RP}$ values are. At $E(B-V) = 0.1$ mag, the values increase from 1.15 at $T_{eff} = 4000$\,K to 1.47 at $T_{eff} = 8000$\,K.
At $T_{eff} = 6000$\,K, the values decreases from 1.30 at $E(B-V) = 0.05$ mag to 1.28 at 0.10 mag.
A very small number of stars whose estimated reddening values are larger than those of the SFD98 map by 0.05 mag are also excluded.

\begin{figure*}
    \centering
    \includegraphics[width=\linewidth]{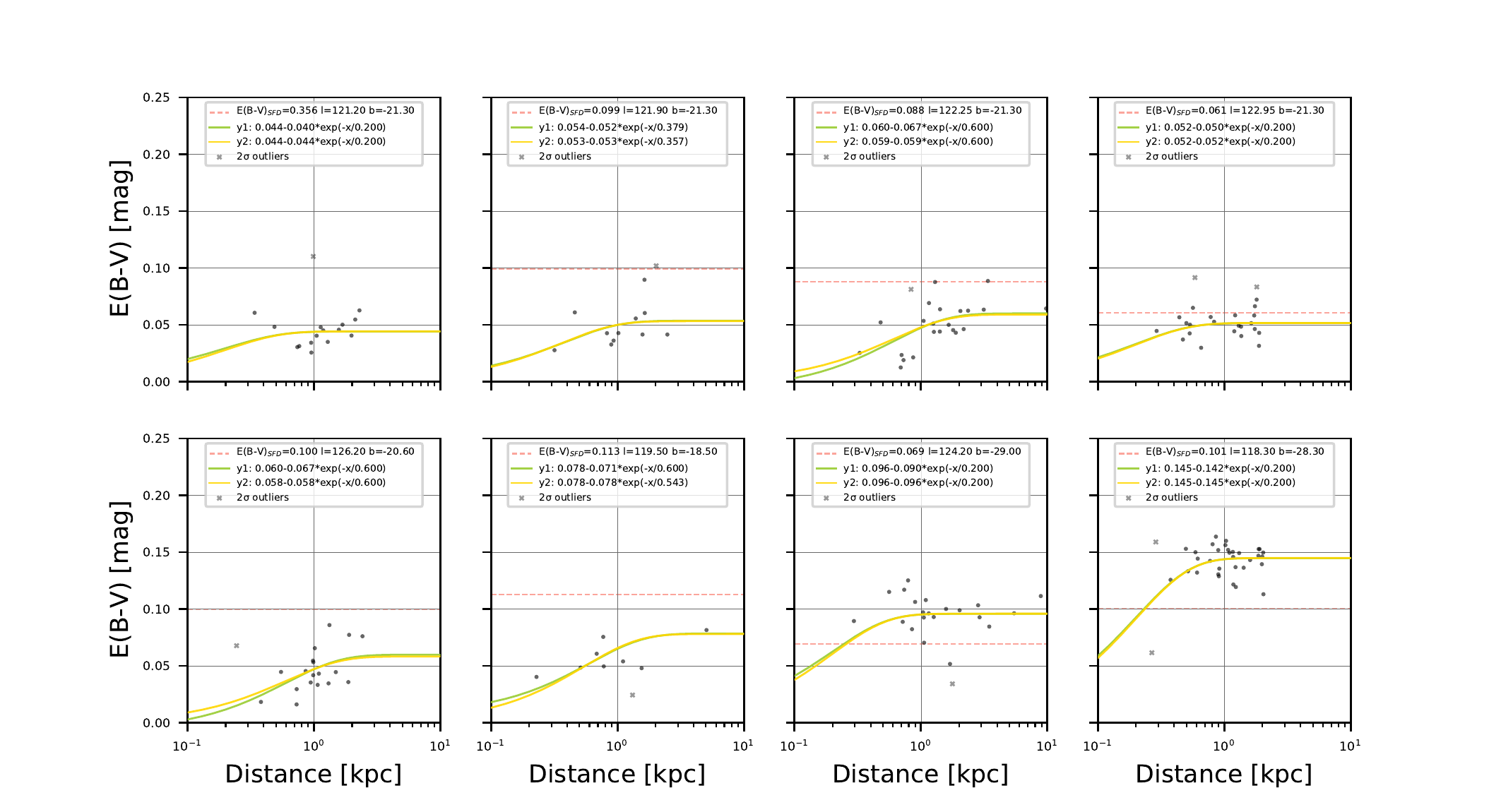}
    \caption{
        \emph{Top panels}: examples of distance-reddening diagram in different sightlines from the M\,31 central region to the outer region.
        \emph{bottom panels}: examples of distance-reddening diagram in sightlines that are far away from M\,31 and show large discrepancies between the SFD98 and LAMOST maps.
        In each panel, black dots are targets used in the fitting, black crosses are outliers,
        green and yellow lines are the fitting results assuming $a\neq b$ and $a=b$,  respectively.
        The E(B-V)$_{SFD}$ values are indicated by red dashed lines for comparison.
    \label{f:EDD}
    }
\end{figure*}

\subsection{Estimates of Galactic foreground reddening for different sightlines}

In previous subsections, we select 193,847 stars within a $25\degree \times 20\degree$ region covering both M\,31 and M\,33 and obtain their reddening values. 
A total $500 \times 400$ sightlines, in intervals of
3' in both the Galactic longitude and latitude directions, are used to construct the 2D Galactic foreground reddening map towards the M\,31 and M\,33 region in this work.
Each sightline contains stars within a small "rectangular" region of $12^\prime \times 12^\prime$. Note that there are common stars between adjacent sightlines.
If there are less than 10 stars for a given sightline, the size is doubled to $24^\prime$ to obtain a reliable result.
If the numbers of stars are still less than 10, the sightlines are masked and not used in the following analysis.
The top left panel of Figure\,1 shows the resulting spatial resolutions. The top right panel shows the histogram distributions of number of stars used for sightlines in the white and gray regions.

Then for each sightline, we use the following function to fit its distance-reddening relation:
\[ y = a - b \cdot e ^{ - \frac{x}{c}} \]
where $x$ is the distance, $y$ is the reddening $E(B-V)$, and $a, b, c$ are three parameters to be constrained.
If we assume a single exponential distribution of dust, then
$a$ and $b$ are equal and represent the total Galactic foreground reddening of a given sightline.
If there is a dust cloud very close to the Galactic disk, then $a$ and $b$ are different,
$a$ still represents the total Galactic foreground reddening, $b$ represents the contribution from the exponential dust disk, and
$a-b$ represents the additional contribution from the dust cloud.
Both scenarios ($a=b$, $a\neq b$) are considered. It's found that
the resulting two $a$ values are very close. Therefore, both $a$ and $b$ parameters are set to be free in this work.

For the $c$ parameter, which is related to the scale height of the dust disk, 
we set lower and upper limits of 0.2 and 0.6 kpc, respectively. The limits correspond to a dust scale height 
between about 70 to 210 pc, within the range of literature values (e.g., Li et al. 2018).
A fitting routine, MPFIT, is used to perform the least-square fit by utilising the Levenberg-Marquardt technique (Markwardt, 2009).
The initial values for $a, b, c$ are adopted to be 0.05, 0.05, and 0.4, respectively. The fitting results are very robust and insensitive to the adopted initial values.
A 2$\sigma$ clipping is also performed during the fitting.

The top panels of Figure\,\ref{f:EDD} show selected examples of fitting results of different sightlines towards M\,31.
In the central region, our fitted values of foreground Galactic extinction are significantly
smaller than the corresponding values from the SFD map, as expected. The differences become
smaller as sightlines get further away. 
The bottom panels of Figure\,\ref{f:EDD} show examples of sightlines that are far away from M\,31 and display large discrepancies between the SFD98 and LAMOST maps.

\section{A two-dimensional foreground dust reddening map} 

\begin{figure*}
    \centering
    \subfigure{
    \includegraphics[width=0.45\linewidth]{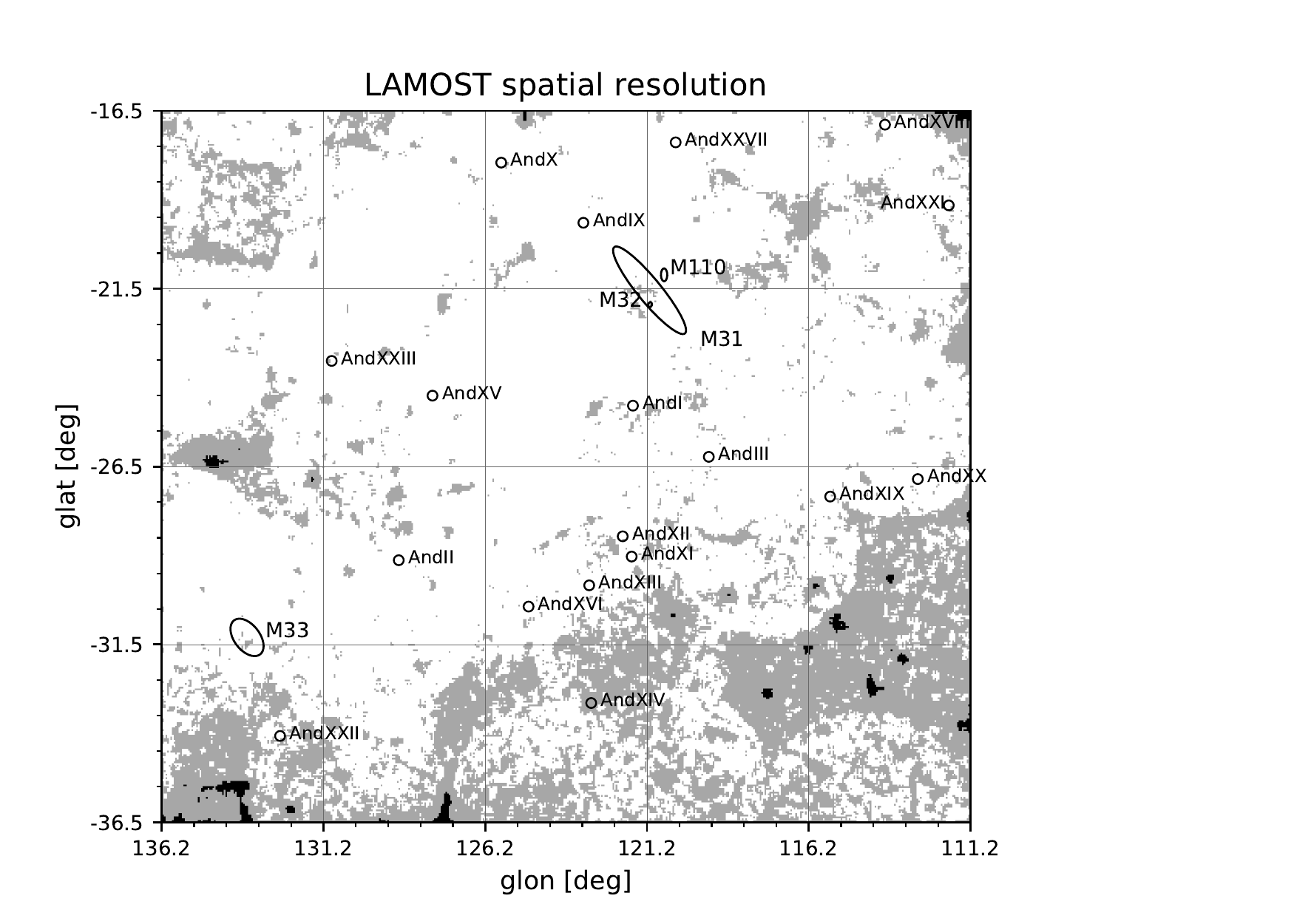}
    } \hspace{-10mm}\vspace{-5mm}
    \subfigure{
    \includegraphics[width=0.45\linewidth]{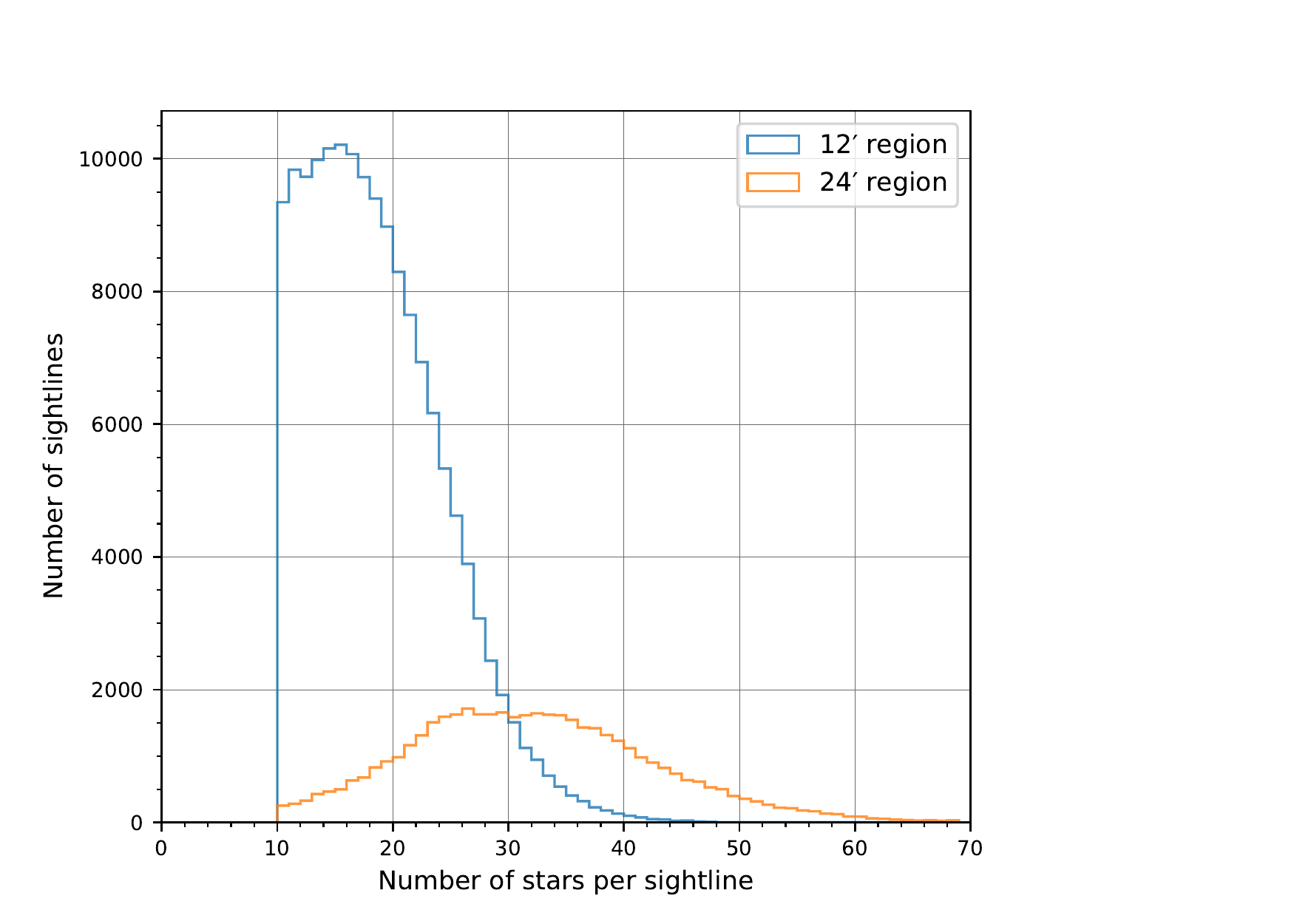}
    } \vspace{-5mm}
    \subfigure{
    \includegraphics[width=0.45\linewidth]{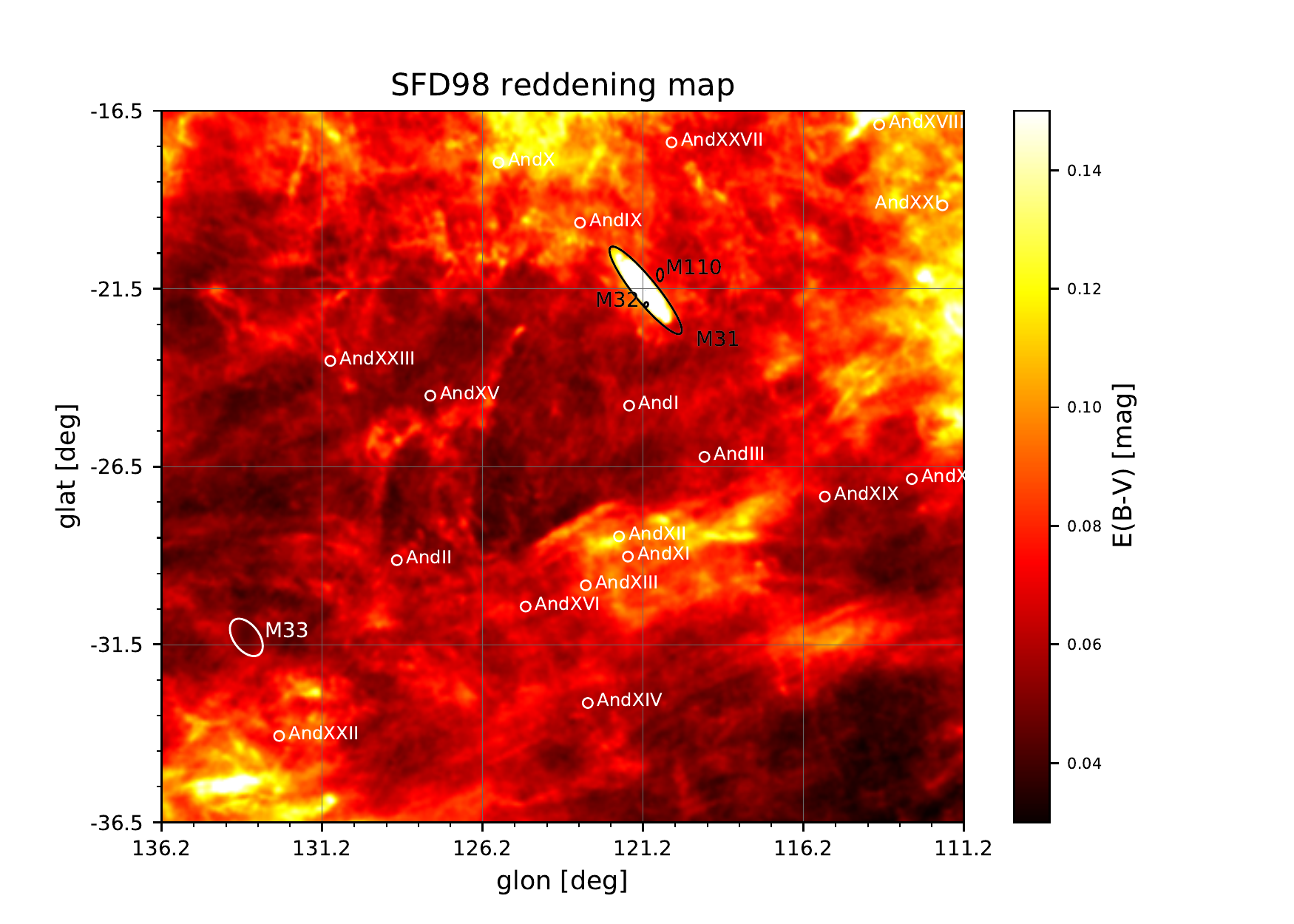}
    } \hspace{-10mm}
    \subfigure{
    \includegraphics[width=0.45\linewidth]{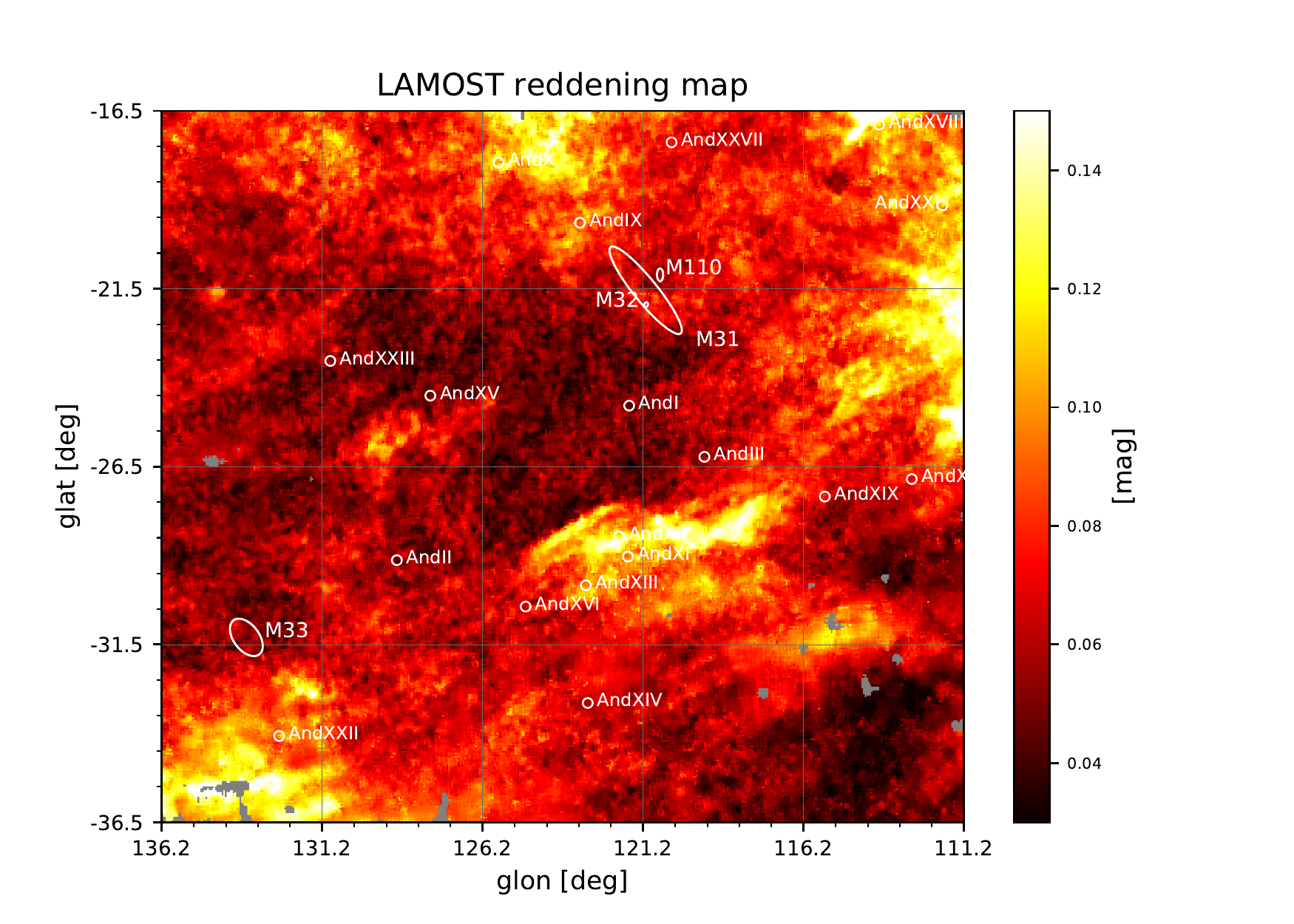}
    } \vfill
    \subfigure{
    \includegraphics[width=0.45\linewidth]{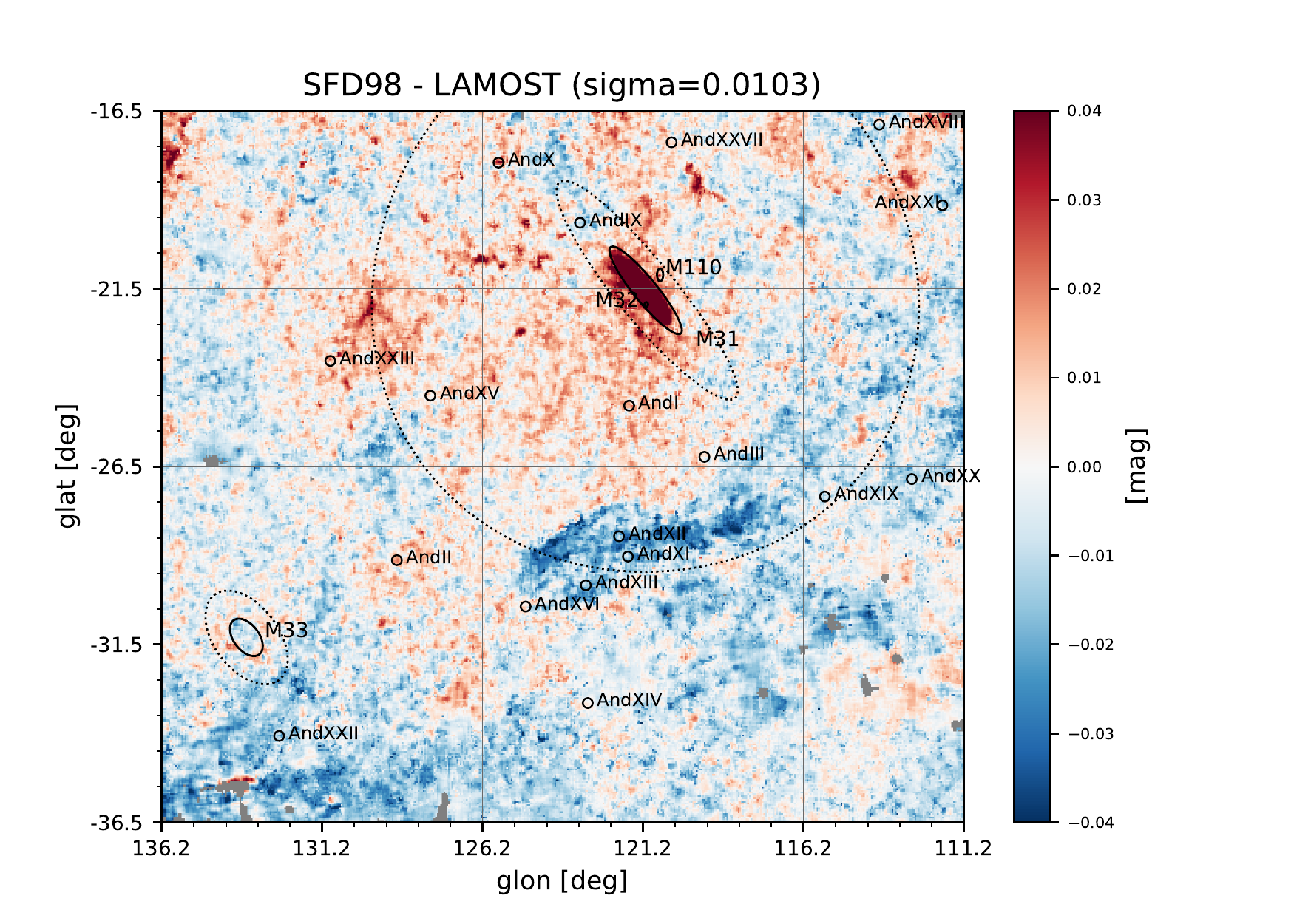}
    } \hspace{-10mm}
    \subfigure{
    \includegraphics[width=0.45\linewidth]{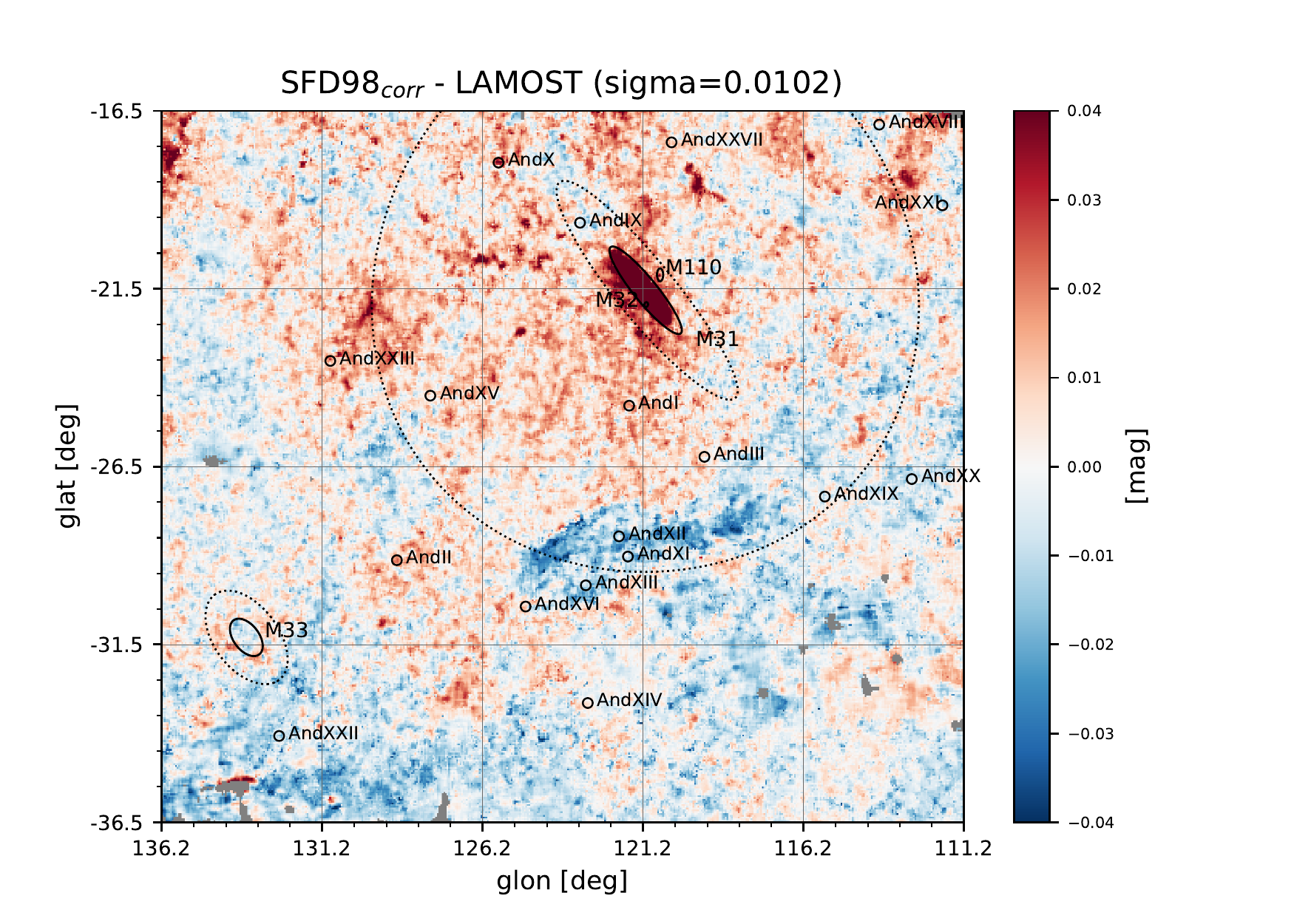}
    } 
    \caption{
        \emph{Top left}:  spatial resolutions (about $12^\prime$ in white regions and $24^\prime$ in gray regions ) of the LAMOST reddening map. Black regions are masked out.
        \emph{Top right}: histogram distributions of number of stars used for sightlines in the white and gray regions.
        \emph{Middle left}:  the SFD98 reddening map.  Dust emission from M\,31 is clearly visible. 
        \emph{Middle right}:  the LAMOST reddening map.  
        \emph{Bottom left}:  the differences between the SFD98 and LAMOST reddening maps. Note the large discrepancies around M\,31.
        \emph{Bottom right}:  the differences between the corrected SFD98 and LAMOST reddening maps, where the SFD98 reddening map is corrected according to its reddening values via a linear relation. 
        In each panel, the solid ellipses mark the optical extent ($R_{25}$) of M\,31, M\,33, and two satellites M\,32 and M\,101. The positions of other satellites of M\,31 are marked by small circles. 
        In the bottom panels, the two dotted ellipses centered on M\,31 and M\,33 represent the extent of their dust disks, 1.5 times larger than their optical extent.
        The large dotted circle centered on M\,31 has a radius of 108 kpc, i.e., 5 $R_{25}$ radii of M\,31. 
    \label{f:maps}
    }
\end{figure*}

So far we have obtained a two-dimensional dust reddening map (LAMOST map hereafter) for the M\,31 and M\,33 region. 
The map has a typical spatial resolution of $12^\prime$, as shown in the top right panel of Figure\,\ref{f:maps}.
Note the SFD98 map has a spatial resolution of $6.^\prime1$, about 2 times higher than that of the LAMOST map.
The LAMOST map, the SFD98 map of the same region, and their differences are displayed in the middle and bottom panels of Figure\,\ref{f:maps}, respectively.
In general, the two maps agree well in most regions. 

The SFD98 map is widely used to perform foreground reddening correction of extragalactic targets. 
Note that a negative offset of 0.003 mag is found for the SFD98 map by comparison with the {\it Planck} result (Planck Collaboration et al. 2014). 
Therefore, 0.003 mag has been added to the SFD98 map throughout this work.
However, due to the difficulties in removing dust emission from very nearby galaxies, such as Magellanic Clouds and M\,31, 
the SFD98 map is not applicable to these galaxies.
As expected and seen in Figure\,\ref{f:maps}, reddening values of the LAMOST map are systematically lower than those of the SFD98 map in regions around M\,31, especially within its optical radius.

Typical values of reddening toward Magellanic Clouds and M\,31 from the SFD98 map are estimated from 
the median dust emission in surrounding annuli: $E(B-V)$ 0.075, 0.037, and 0.062 mag for the Large Magellanic Cloud, Small Magellanic Cloud, and M\,31, respectively (Schlegel et al. 1998). 
Given the large field of view of these galaxies, Galactic foreground reddening corrections using the aforementioned values probably suffer large uncertainties.
The LAMOST map shows that there are Galactic foreground "cirrus" clouds in front of the M\,31 galaxy. Within the optical radius, 
the median value of reddening is 0.069 mag, the peak-to-peak value is 0.035--0.097 mag, the dispersion is 0.011 mag.  
The precise Galactic foreground reddening map obtained in this work is also very crucial in order to 
1) measure precise extinction curves across different regions of M\,31 to study their variations with different environments (Bianchi et al. 1996; Dong et al. 2014; Clayton et al. 2015) 
and then 2) perform better reddening correction of M\,31 targets to 
reveal the stellar populations, the structure of M\,31 using photometry of large numbers of individual stars (e.g., Dalcanton et al. 2012).
Note that dust emission in the central part of M\,110 is also detected in the bottom left panel of Figure\,\ref{f:maps}, consistent with the fact that
M\,110 contains a population of young blue stars at its center. 

\begin{figure}
    \centering
    \includegraphics[width=\linewidth]{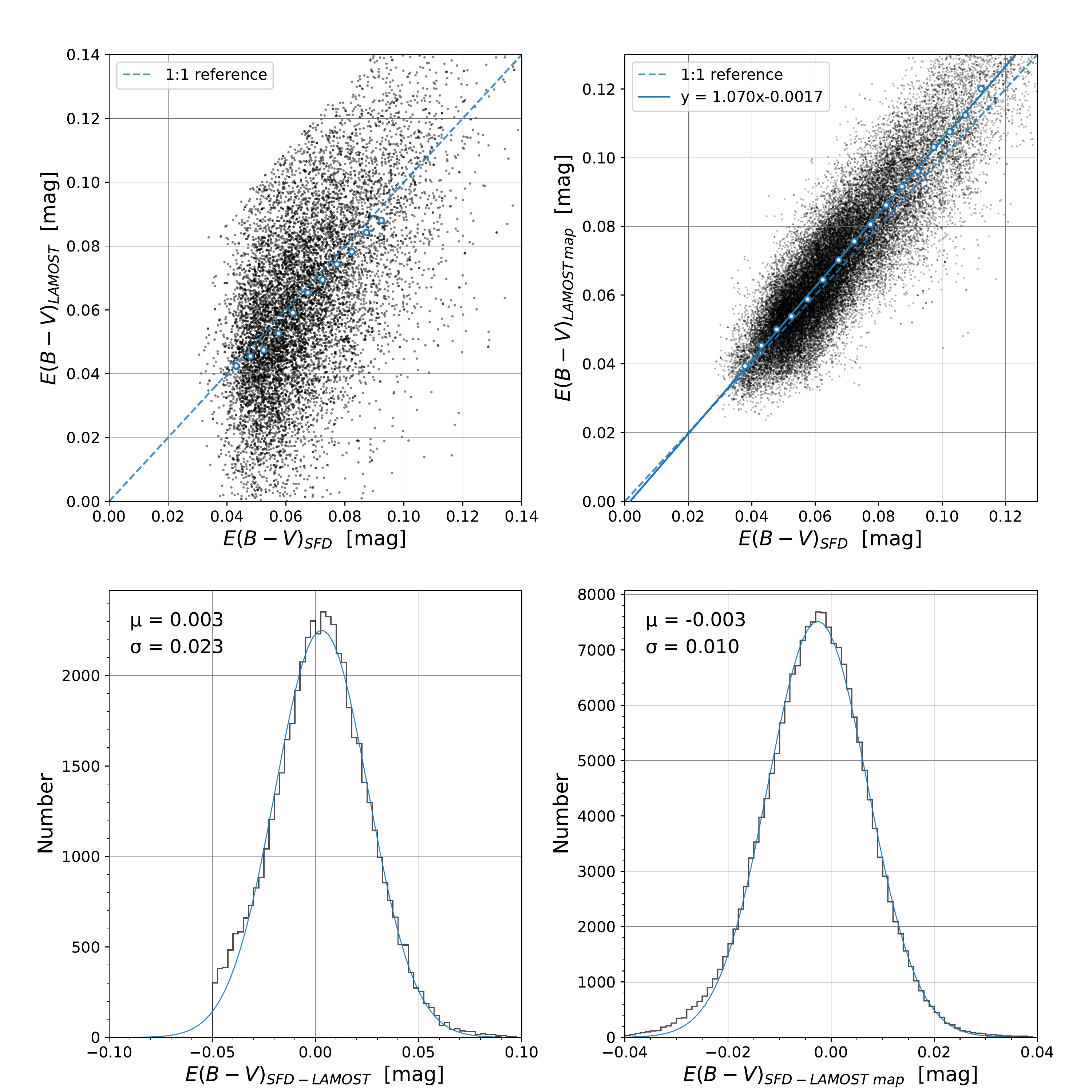}
    \caption{
        \emph{Top left}: comparison of LAMOST reddening estimates with those from the SFD98 map for selected individual stars.
        \emph{Top right}: comparison of the LAMOST map with the SFD98 map for selected regions.
        The points in the top two panels are divided into different bins and the white circles indicate the median values. The blue dashed lines indicate lines of equality, and the blue solid line in the top right panel
        indicate the linear fitting result to the white circles.
        \emph{Bottom left}: histogram distribution and gaussian fit of $E(B-V)_{SFD-LAMOST}$  for the selected stars.
        \emph{Bottom right}: histogram distribution and gaussian fit of  $E(B-V)_{SFD-LAMOST}$  for the selected regions.
    \label{f:comparison}
    }
\end{figure}

To check the precision of the LAMOST map, we first select stars that are far above the Galactic plane ($|Z| > 1.2$ kpc) and far away from M\,31 
($gl > 124.2^\circ$ or $gl < 118.2^\circ$ or $gb > -18.5^\circ$ or $gb < -24.5^\circ$). 
The left panels of Figure\,\ref{f:comparison} plot comparisons of LAMOST reddening estimates with those from the SFD98 map for the selected stars.
The two measurements agree well at a dispersion of 0.023 mag, suggesting that the reddening estimates have a typical precision of 0.023 mag for individual stars. 
The precision depends on SNRs of LAMOST stars. For stars of SNRs lower than 20, the precision decreases to 0.026 mag. 
There are a tail of stars that have systematically lower LAMOST reddening than SFD reddening. Those stars have typically lower SNRs.
Note the LAMOST estimates are systematically lower by a very small number of 0.003 mag. 
The right panels of Figure\,\ref{f:comparison} compare the LAMOST map with the SFD98 map. The same region centered on M\,31 
($118.2^\circ \le gl \le 124.2^\circ$, $-24.5^\circ \le gb \le -18.5^\circ$) is also excluded for comparison. 
The dispersion is only 0.01 mag, indicating that the LAMOST map has a high precision of 0.01 mag. However, the SFD98 map is systematically lower by 0.003 mag, and 
the number shows a clear dependence on reddening. The discrepancies are larger at regions of higher extinction or colder dust temperature. 
consistent with the findings of Sun et al. (in preparation) that the SFD98 map suffers systematics that depends on dust reddening, dust temperature and positions.

\section{Dust distribution in the outskirts of M\,31 and M\,33}

\begin{figure*}
    \centering
    \includegraphics[width=\linewidth]{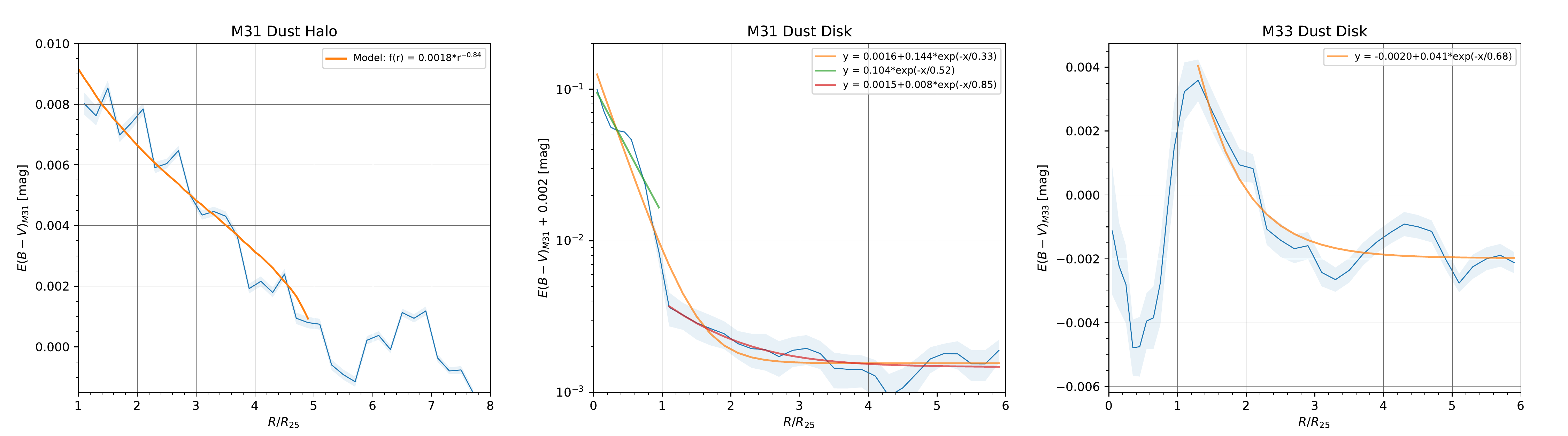}
    \caption{
        \emph{Left}: observed (blue) and modeled (orange) radial trend of dust extinction in the M\,31 halo. A simple spherically symmetric power-law distribution of dust is assumed.
        \emph{Middle}: observed (blue) and modeled radial trends of dust extinction in the M\,31 disk, seen as if it is edge-on. 
        An offset of 0.002 mag is added to the Y-axis. Note the prominent reddening jump at its optical radius.
                         The yellow line represents a single exponential fit for the dust disk, 
                         and the green and red lines are exponential fits for dust disks within and outside its optical radius, respectively. 
        \emph{Right}: observed (blue) and modeled (orange) radial trend of dust extinction in the M\,33 disk, seen as if it is edge-on. An exponential disk model is adopted. 
    \label{f:disk}
    }
\end{figure*}

To study dust distribution in the outskirts of M\,31 and M\,33, we first correct for the reddening dependent systematics of the SFD98 map for regions outside the 
optical radius of M\,31 using the linear formula
$E(B-V)_{corr} = 1.070 * E(B-V) - 0.0017$, as indicated in the top right panel of Figure\,3. 
The updated residual map is shown in the bottom right panel of Figure\,\ref{f:maps}. 
To avoid possible uncertainties in the correction, only regions of $E(B-V)_{SFD}\, \textless \,0.09$ mag are used in the following analysis. 
Radial trends are then studied by averaging in annuli centered on the two galaxies. Their errors are also estimated by their standard deviations divided by the square root of the number of pixels. 

Clumpy and filamentary structures surrounding M\,31 point to the possible existence of dust (emission) in the CGM of M\,31. 
Circular annuli centered on M\,31 are firstly used to study dust distribution in the halo of M\,31.
To avoid possible contributions from the M\,31 dust disk, only regions outside its 4 $R_{25}$ radii are used. 
The result is shown in the left panel of Figure 4. It can be seen that dust is detected in the M\,31 halo up to a distance of $\sim$ 5 $R_{25}$ radii, i.e., 108 kpc. 
Reddening caused by dust in the M\,31 halo can be as high as 0.01 mag in the center region and drops to below 0.001 mag at a projected distance of $\sim$ 5 $R_{25}$.
The numbers agree very well with those of M{\'e}nard et al. (2010), who find that a background source is typically
reddened from a $z \sim 0.3$ galaxy by about 0.01 mag at a separation of 20 kpc , and by about 0.001 mag at 100 kpc.
A simple spherically symmetric power-law distribution of dust in the M\,31 halo, with galactocentric distance r between 1.1 -- 5 $R_{25}$ radii, is adopted to fit the observed curve.
The resulting curve in over-plotted in orange and agrees well with the observed one. 
A power law index of $-0.84$ is obtained, suggesting higher dust densities in the inner halo but more dust in the outer halo.
Without correction for the reddening dependent systematics of the SFD98 map, the resulting curve is systematically lower by about 2.5 mmag, varying slightly with radius.
In such case, the dust halo can still be detected up to a distance of $\sim$ 4 $R_{25}$ radii. Beyond 4 $R_{25}$ radii, negative reddening values are detected, suggesting that 
the correction is necessary.

Elliptical annuli centered on M\,31 are also used to study dust distribution in the disk of M\,31. 
Possible contaminations from the M\,31 halo are subtracted according to the observed curve in the left panel of Figure\,4. 
For regions within 1.1 $R_{25}$ radii, 0.008 mag is adopted.
The project effect has also been corrected by multiplying $cos(77.8^\circ)=0.21$, where 77.8$^\circ$ is the inclination of the M\,31 disk.
The middle panel of Figure 4 plot the reddening of the M\,31 dust disk as a function of galactocentric distance, seen as if it is edge-on.
Dust in the M\,31 disk is found to extend out to about 2.5 times its optical radius, whose distribution is 
consistent with either an exponential of scale length of 7.2 kpc or an exponential of scale length of 11.1 kpc within its optical radius and another one of 18.3 kpc beyond its optical radius. 
Our result is consistent with Tempel et al. (2010), who find a dust disk scale length of 9.8 kpc within its optical radius, about 1.8 times the stellar scale length.
In the latter case, we note a prominent reddening jump at its optical radius. 
If without correction for the reddening dependent systematics of the SFD98 map, the resulting curve for the M\,31 dust disk is hardly changed.
Note that to make the plot in the logarithmic scale, a constant of 0.002 mag is added to make sure all the data points are positive.
At $R > 2.5 R_{25}$,  negative reddening of the M\,31 dust disk is detected. This is probably caused by an over-subtraction of the halo component. 
As discussed in the coming section, there is a possible lack of dust in the halo of M\,31 along its major axis direction.


Elliptical annuli are also used to study dust distribution in the outskirt of M\,33 disk.
The project effect is corrected as in M\,31.
Note that the dust emission within one $R_{25}$ radius of M\,33 was removed from the SFD98 map. 
As can be seen from the middle left panel of Figure\,2, the reddening values within its $R_{25}$ radius are systematically lower 
than those of adjacent regions, suggesting that an over-subtraction probably happened.
Outside its $R_{25}$ radius, a significant dust signal is detected up to $\sim$ 2.5 $R_{25}$ radii and can be well described by an exponential disk 
with a scale length of 0.68 $R_{25}$, i.e., 24.1$^\prime$ and 5.7 kpc.
This number is very close to the scale length of its extended outer stellar disk (25.4$^\prime$, 6.0 kpc, Grossi et al. 2011), 1.5 times larger than 
its stellar scale length in optical (9.6$^\prime$, van den Bergh 1991).
If without correction for the reddening dependent systematics of the SFD98 map, the resulting curve for the M\,33 dust disk is systematically lower by about 1.5 mmag, 
but the profile is unchanged.
At $R > 2 R_{25}$,  negative reddening of the M\,33 dust disk is detected. This is probably caused by systematic errors in the correction of the SFD98 map,
particularly for the blue region in the lower right of M\,33.
The SFD98 map suffers systematics that depends on dust reddening, dust temperature and positions, 
and we only make a simple reddening dependent linear correction in this work.

\section{DISCUSSION}

We find that M\,31 has a large and dusty halo. This result 
is consistent with the work of Lehner et al. (2015, 2020) who show the presence of an extended and massive CGM around M\,31 via absorption line studies of dozens of quasars.
The result also agrees well with M{\'e}nard et al. (2010), who find strong evidence for the existence of a diffuse component of dust in galactic haloes, extending from 
20 kpc to several Mpc. The projected reddening profiles from 20 kpc to 100 kpc are also similar, as mentioned in the previous section. 
Using cosmological hydrodynamical simulations, P{\'e}roux et al. (2020)  recently report a strong dependence of gas mass flow rates and gas metallicity of CGM 
on azimuthal angle with respect to its central galaxy: outflows are more favored along the galaxy minor axis and tends to have higher metallicity than inflows. 
Therefore, one would expect less dust along the galaxy major axis direction.
We note a possible lack of dust in the halo of M\,31 along its major axis direction, thus may providing a direct observational evidence for the findings of P{\'e}roux et al.

M{\'e}nard et al. estimate that the dust mass in the halo of Milky Way-like galaxies ($\sim 5 \times 10^7 M_\odot$ ) is comparable to that commonly found in galactic discs (Draine et al. 2007).
Assuming simply uniform dust properties across the disk and halo of M\,31, we have estimated the mass fractions of dust in its disk (0 -- 2.5 $R_{25}$)  and 
halo (0 -- 5 $R_{25}$) according to their reddening profiles (Figure\,4).
We find that roughly 75 percent of dust is in the halo and only 25 percent is in the disk, and only 1.5 percent in its outer disk (1 -- 2.5 $R_{25}$). 
Based on a physical dust model, Draine et al. (2014) have estimated a total dust mass of $5.4 \times 10^7 M_\odot$ in the disk of M\,31 out to a distance of 25 kpc.
Following the method of M{\'e}nard et al. (see their Section 5.1), we estimate a total dust mass in the halo of M\,31 to be $\sim 1.8 \times 10^8 M_\odot$, 
assuming an SMC-type dust and $R_{\it V} = 4.9$. The above estimates suggest that about three-quarters of dust is in the halo of M\,31. 
If without correction for the reddening dependent systematics of the SFD98 map, the total dust mass in the  halo of M\,31 (0 -- 4 $R_{25}$) 
decreases to  $\sim 0.77 \times 10^8 M_\odot$, the dust mass in the disk is unchanged. In such case,  about 55 per cent of dust is in the halo of M\,31.

In the above analysis, we have ignored the contributions of dust in the Galactic halo to the reddening of M\,31 halo. 
If there are distant dust clouds in the Galactic halo that are not probed by our sampling stars, the resulting LAMOST reddening map will be under-estimated, 
and reddening from the M\,31 halo will be over-estimated. However, considering the rapidly decreasing reddening profile for the M\,31 halo, 
the effects of dust in the Galactic halo should be weak.

A number of radio observations have revealed an M\,31 neutral gaseous halo extending from its disk to at least halfway to M\,33 (Braun \& Thilker 2004; Lockman et al. 2012; Kerp et al. 2016). 
About half of gas in its gaseous halo is in an extended, diffuse component, with another half composed of clouds that are likely to be fuel for future star formation in M\,31 and M\,33 (Wolfe et al. 2013).
Further explorations of properties of dust in the outskirts and CGM of M\,31 and M\,33 and associations with gas are very promising to constrain galactic outflows and recycling, 
one of the key ingredients for understanding galaxy evolution. 

\section{SUMMARY}

In this work, using 193,847 LAMOST stars with precise reddening estimates and parallaxes, 
we have constructed a large two-dimensional foreground dust reddening map towards the M\,31 and M\,33 region ($111.2^\circ \le gl \le 136.2^\circ$, $-36.5^\circ \le gb \le -16.5^\circ$),
at a typical spatial resolution of about $12^\prime$. 
The map agrees well with the SFD98 map in most regions and has a typical precision of 0.01 mag.
The map shows the complex structure of dust clouds towards the M\,31, suggesting that the map should be used 
for precise foreground reddening corrections in studies such as measuring variations of extinction curves across different regions of M\,31.

By carefully removing the foreground reddening from the SFD98 map, 
the distribution of dust in the outskirts of M\,31 and M\,33 are revealed in great details to a very large distance.
We find that a significant amount of dust in clumpy and filamentary structures exists in the halo of M\,31, out to a distance of over 100 kpc.
Reddening caused by dust in the M\,31 halo decreases from about 0.01 mag at the center to 0.001 mag at a projected distance of $\sim$ 5 $R_{25}$.
Dust in the M\,31 disk extends out to about 2.5 times its optical radius, whose distribution can be described by either an exponential disk of scale length of 7.2 kpc or 
two disks with scale length of 11.1 kpc within its optical radius and 18.3 kpc beyond its optical radius. 
Dust in the disk of M\,33 is also found to extend out to about 2.5 times its optical radius, beyond one optical radius, its distribution is 
consistent with an exponential disk of scale length of 5.6 kpc.
Our results combined with future observations will provide new clues on the distributions, properties, and cycling of dust in spiral galaxies.

\begin{acknowledgements}
We acknowledge the anonymous referee for his/her valuable comments that improve the quality of this paper.
We acknowledge very helpful discussions with Xiaowei Liu, Biwei Jiang, Jiang Gao, Bingqiu Chen, Maosheng Xiang, and Yang Huang.
This work is supported by the National Key Basic R\&D Program of China via 2019YFA0405500, National Natural Science Foundation of China through the project NSFC 11603002, 
 and Beijing Normal University grant No. 310232102.
Guoshoujing Telescope (the Large Sky Area Multi-Object Fiber Spectroscopic Telescope LAMOST) is a National Major Scientific Project built by the Chinese Academy of Sciences. Funding for the project has been provided by the National Development and Reform Commission. LAMOST is operated and managed by the National Astronomical Observatories, Chinese Academy of Sciences.
This work has made use of data from the European Space Agency (ESA) mission {\it Gaia} (https://www.cosmos.esa.int/gaia), processed by the {\it Gaia} Data Processing and Analysis 
Consortium (DPAC, https://www.cosmos.esa.int/
web/gaia/dpac/ consortium). Funding for the DPAC has been provided by national institutions, in particular the institutions participating in the {\it Gaia} Multilateral Agreement.
This research has made use of the SIMBAD database,
operated at CDS, Strasbourg, France
\end{acknowledgements}


\end{document}